# AI Healthcare Chatbots as Information Infrastructure: A Large-Scale Study of User-Reported Breakdowns

**Hassan, Muhammad** University of Illinois Urbana Champaign, USA | mhassa42@illinois.edu
**Yener, Ramazan** University of Illinois Urbana Champaign, USA | ryener2@illinois.edu
**Gumusel, Ece** Rutgers University, USA | ece.g@rutgers.edu
**Bashir, Masooda** University of Illinois Urbana Champaign, USA | mnb@illinois.edu

## ABSTRACT

AI healthcare chatbots are increasingly used to support health information seeking and self-management, yet their performance and impact on users remains to be studied. This study examines over 15,000 user reviews from 59 AI healthcare chatbot apps to explore how these systems function in everyday informational and emotional contexts. Topic modeling and interpretive analysis identify three recurring breakdowns: access barriers and service unreliability, user experience and interaction quality, and billing and customer support issues. Privacy and security concerns are associated with the most negative experiences. By framing AI healthcare chatbots as information infrastructures, our findings highlight how failures in access, usability, and trust affect users, offering actionable insights for designers, policymakers, and information professionals aiming to improve digital health systems.

## KEYWORDS

AI Health; Digital Wellbeing; User Experience; Information Infrastructures.

## INTRODUCTION

AI-enabled healthcare chatbots have become increasingly embedded in digital health ecosystems, shaping how individuals access and engage with health-related information. Recent Pew Research Center data indicate that 22% of Americans report using AI chatbots for health-related reasons, and younger adults are more likely than older peers to turn to AI chatbots and social media for healthcare needs (Pew Research Center, 2026). Delivered primarily through mobile applications, these systems leverage smartphone affordance, such as push notifications, sensors, and always-available interfaces, to provide continuous and on-demand access to informational resources, self-management guidance, and conversational support that extends beyond formal clinical settings (Laranjo et al., 2018). Applications such as Wysa, Youper, and Woebot illustrate how conversational agents are positioned as everyday tools for managing mood, symptoms, and well-being (Inkster et al., 2018). As these systems become integrated within users' ongoing information practices, examining how individuals experience them in situ, particularly when users' expectations are unmet, becomes a concern. This perspective highlights the need to understand user concerns and dissatisfaction not as isolated failure, but as a meaningful signal of how such systems operate within lived contexts of everyday users.

Conceptualizing AI healthcare chatbots as information infrastructures highlights their role in mediating access to knowledge, support, and services across intersecting domains (Star & Bowker, 1999). Infrastructures often become most visible when they break down, when access is blocked, interactions fail, or institutional arrangements such as payment and support become unclear (Star & Ruhleder, 1996. Health chatbots operate across contextual boundaries: between clinical and everyday environments, between institutional healthcare systems and commercial platforms, and between personal experiences and distributed informational resources. Attending to users' negative experiences therefore offers an analytical lens for examining how these infrastructures function in practice and where they fall short. Such an approach enables a closer understanding of how infrastructural conditions shape, and at times constrain users' experiences of informational support and well-being. Examining these user experiences offers insight into how these systems function in practice and where they fall short.

Existing research, such as Fitzpatrick et al. (2017), has examined AI-incorporated healthcare technologies, including conversational agents designed for symptom checking, behavioral coaching, and emotional support. Prior studies have also identified concerns related to information privacy and governance in AI-enabled health applications (Williamson & Prybutok, 2024; Reddy et al., 2020). App-store reviews represent a significant yet underexamined site of information practice, where users actively interpret, evaluate, and publicly document their experiences. These reviews form a rich corpus through which user expectations, frustrations, and convictions about health technologies become visible, offering insight into how users make sense of these systems in everyday use. However, there are no large-scale analyses of the app users' existing challenges in these chatbots, particularly in relation to *how AI healthcare chatbot users articulate concerns about access, interaction quality, billing arrangements, and data practices in mobile applications*.

In this paper, we investigate user feedback reviews for AI healthcare chatbot applications they use in everyday life. We collected the publicly available reviews from Google Play Store and the Apple App Store in late 2025. Drawing on over 15,000 reviews across 38 Android and 21 iOS applications, we employ topic modeling to identify recurring patterns in user complaints and interpretive analysis to organize these patterns into three higher-level categories of concern. In addition, we apply keyword-based search to identify reviews that explicitly reference privacy, security, or data-handling issues, treating these as an additional lens on user-expressed perceptions of risk and trust. Our analysis is guided by the following questions

- How do users describe barriers to accessing and reliably using AI healthcare chatbots?
- How do users characterize interaction quality and perceived usefulness?
- How do users experience trust, fairness, support structures, and explicit privacy/security concerns?

Together, these questions structure an inquiry into how users experience issues across AI health chatbot systems.

This study makes three contributions to information science research. First, it offers a large-scale, empirically grounded account of how users articulate dissatisfaction with AI healthcare chatbots in naturally occurring settings, complementing prior work based on smaller samples or clinical evaluations (Miner et al., 2016). Second, it identifies three interrelated categories of information problems, access and reliability, interaction quality, and billing and support structures, that shape users' negative experiences and reflect broader challenges in the design and governance of information systems (van Dijck et al., 2018). Third, it demonstrates that explicit references to privacy, security, and data handling, while relatively infrequent, are associated with more severe negative ratings, indicating that such concerns emerge as markers of heightened distrust rather than routine critique (Mittelstadt et al., 2016). Taken together, these findings position AI healthcare chatbots within a broader landscape of evolving health information infrastructures, where questions of usability, trust, and governance remain central to how informational support is distributed and experienced in everyday life.

## BACKGROUND AND RELATED WORK

### AI Healthcare Chatbots as Information Systems

AI-enabled healthcare chatbots have evolved into widely used digital wellness tools designed to support health ecosystems, reshaping how individuals access and engage with health-related information. These systems, ranging from rule-based conversational agents to more recent machine learning and large language models (LLM)-driven tools, are designed to support symptom checking, self-management, and emotional or mental health support through conversational interaction (Laranjo et al., 2018; Inkster et al., 2018). Prior research has demonstrated their potential to extend access to health information and provide scalable forms of support, particularly in contexts where traditional care is limited or inaccessible (Fitzpatrick et al., 2017).

These systems are sociotechnical arrangements that mediate access to knowledge, support, and services across clinical and everyday settings (Ocepek, 2018). Users engage with chatbots not only to retrieve information but also to interpret symptoms, reflect on personal experiences, and seek emotional reassurance. This positions AI healthcare chatbots within broader frameworks of everyday information seeking and use, where informational and affective needs are often intertwined (Savolainen, 2005; Wardle et al., 2025; Liu et al., 2025). In this sense, chatbots function not merely as tools, but as intermediaries that shape how health information is encountered, interpreted, and applied in practice, both for intended health support and more exploratory use as discussed by Chen et al, (2024).

At the same time, a growing body of work has raised concerns about the reliability, safety, and limitations of these systems. Studies have identified issues such as inaccurate or incomplete responses, lack of clinical grounding, and inconsistent conversational behavior (Miner at al., 2016; Bickmore & Giorgino, 2006). In health contexts, concerns extend further to include the potential for inappropriate responses, failure to recognize crisis situations, and the risk of users placing undue trust in automated systems (Vaidyam et al., 2019). These challenges highlight the need to examine not only the technical capabilities of chatbots, but also how they perform in real-world use, particularly in relation to user expectations and well-being.

### Trust, Privacy, and Data Practices in Digital Health Systems

Trust is a central concern in the use of digital health technologies, particularly when systems handle sensitive personal and health-related information. Work in ethics and law has shown that perceptions of AI-based systems depend not only on technical performance but also on how clearly data practices and responsibilities are defined. Mittelstadt et al. (2016), for example, identify key ethical challenges of algorithmic decision support, including opacity and unclear accountability in data use, while Veale et al. (2018) examine how data-protection and non-discrimination law apply to automated decision-making, highlighting uncertainties around explanation and redress. Williamson and Prybutok (2024) analyse user responses to AI-enabled health applications, showing that privacy, security, and transparency concerns are closely tied to whether users perceive such tools as trustworthy. Together, this work underscores that data practices are not a background detail but a visible component of how digital health systems are evaluated.

Empirical studies of concrete systems deepen these concerns by showing how users face both limited control and partial information even where formal policies exist. Horstmann et al. (2024) find that people struggle to understand what data applications collect and share, complicating informed consent, while Winter and Davidson (2019) show that patients often lack clarity about how their data flow across platforms, raising questions of accountability and governance when multiple actors are involved. At the same time, existing work indicates that people do not always explicitly articulate privacy or security concerns in everyday use: Ischen et al. (2020) suggest that privacy worries may remain latent until a service behaves unexpectedly, and Gumusel (2025) shows how users express unease indirectly, through discomfort or distrust rather than explicit privacy language, underscoring that the absence of privacy terminology in app reviews does not necessarily imply the absence of privacy-related discomfort.

Literature on AI health chatbots highlights both data sensitivity and regulatory gaps in consumer-facing tools. Haupt and Marks (2023) show that mental-health chatbots maybe between clinical regulation and consumer protection, creating ambiguous protections around data handling and safety, while Hassan and Bashir (2023) identify mismatches between apps stated privacy policies and their data flows. Unlike clinical systems governed by established health-data frameworks, chatbots embedded in commercial apps operate under less clearly defined standards, so worries about privacy and data use often intersect with broader issues of trust, including payment practices and service reliability. At the same time, studies of conversational agents underscore limitations in reliability and safety: Miner et al. (2016) and Bickmore and Giorgino (2006) document inaccuracies, weak clinical grounding, and inconsistent dialogue, and Vaidyam et al. (2019) show that failures to recognise or appropriately respond to crisis situations can pose substantial risk. Together, this work suggests the need to evaluate not only how chatbots manage data, but also how they perform in real-world use relative to user expectations and well-being.

Building on this literature, the present study investigates users' experience with AI healthcare chatbots and investigates security-, privacy-, and data-handling-related concerns (SPR) as a specific dimension of user experience. We examine how such concerns are explicitly articulated within reviews and how they relate to broader patterns of dissatisfaction, including breakdowns in access, interaction quality, and financial trust. This approach allows us to situate SPR-related complaints within a wider landscape of infrastructural issues, rather than treating them as an isolated problem of compliance or policy.

### User Reviews as Information Experience and Evaluation

User-generated reviews provide a valuable lens for understanding how digital systems are experienced and evaluated in everyday contexts. App-store reviews capture spontaneous and situated accounts of user interaction, often written at moments when expectations are unmet or experiences are especially salient. Prior work has used such reviews to investigate usability issues, trust, and perceived quality across a range of applications, including health technologies. For example, Pagano and Maalej (2013) analyse app-store reviews to surface recurring feature requests and quality problems, demonstrating how reviews can guide systematic identification of design issues. Fu et al. (2013) focus specifically on mobile applications, showing how user feedback reveals concerns about safety, usability, and appropriateness of content. User reviews often combine descriptive accounts of system use with evaluative judgments, highlighting the criteria users apply when assessing usefulness, reliability, and trustworthiness (Savolainen, 2007; Talja & Hansen, 2006). In this sense, they serve not only as feedback to developers but also as public narratives that shape collective understanding of applications and circulate expectations about how these apps should behave. Together, these studies establish online discussion, such as user reviews, as a methodologically robust source for studying user experience and for identifying patterns that may not be visible through controlled evaluations alone.

Studies of health technologies have extended this perspective to a variety of online settings. Andalibi et al. (2018), for instance, examine how people disclose and seek support around mental health on social media, while Andalibi (2020) analyses how platform features shape the ways users talk about sensitive experiences. This work shows that user-generated content offers insight into how people negotiate stigma, seek help, and evaluate sociotechnical systems in context. However, much of the existing research focuses on specific platforms or comparatively small samples, which limits the ability to identify broader, cross-system patterns in how users articulate dissatisfaction with access, interaction quality, and service structures.

By focusing on app-store reviews of AI healthcare chatbots, this study builds on and extends these strands of work. Evaluations of user dissatisfaction through reviews make explicit the points at which systems fail to meet user expectations, offering analytically productive moments where infrastructural conditions and trust relations become visible. In combination with prior research on reliability, safety, and data practices in digital health systems, treating user reviews as information practice allows us to connect concerns about access, interaction quality, payment and support practices, and privacy to users' own accounts of how these chatbots function, or fail to function, in everyday life. This approach aligns with a broader tradition that treats failure and friction as revealing of underlying infrastructural conditions (Star & Ruhleder, 1996), and it provides the methodological grounding for our use of large-scale review analysis in the remainder of the paper.

## METHODOLOGY

This study combines large-scale corpus construction with computational modelling and interpretive analysis. In this section, we describe how we identified AI-enabled healthcare chatbot applications, collected and filtered app-store reviews, and analyzed the resulting corpus.

### Step 1: Data Collection

To construct an empirical corpus, we systematically identified AI-enabled healthcare chatbot applications from the Google Play Store (Android) and the Apple App Store (iOS). We did not apply any regional filters during collection, so the resulting apps and reviews are likely to reflect a global user base rather than a regional context. We conducted keyword-based searches using terms such as "AI health," "AI healthcare," and "healthcare chatbot," combining health-related and AI-oriented terms to capture a broad range of applications. Search results were aggregated and deduplicated based on app name and developer.

We applied two inclusion criteria to ensure analytical relevance. First, the application had to focus on health, as indicated by its store category, title, or description. Second, it had to provide AI-based conversational or chatbot functionality, whether implemented via rule-based dialogue, proprietary machine-learning models, or LLM-based components embedded through mobile SDKs. Two researchers independently coded each app as "include" or "exclude" using publicly available store materials, including descriptions, screenshots, and interface cues. Inter-rater reliability was assessed using Cohen's kappa ($\kappa = 0.6773$), indicating substantial agreement, and the discrepancies were resolved through discussion until consensus was reached. This process yielded a final sample of 38 Android applications and 21 iOS applications, including 18 applications available on both platforms.

### Step 2: User-reviews corpus and preprocessing

We next constructed a large-scale corpus of user reviews associated with the selected applications. Reviews were collected using established Python-based scraping tools (e.g., *google-play-scraper*, *app_store_scraper*) (JoMingyu, 2024; Lim, 2020), consistent with prior research on app-store data as a source of user feedback (Nissen et al., 2024). This process yielded 264,310 reviews across both platforms, each accompanied by metadata such as platform, app identifier, rating, and timestamp where available. The reviews span approximately fourteen years of activity, from June 2011 to October 2025.

To maintain linguistic consistency, we restricted the dataset to English-language reviews using automated language detection (Danilk, 2016), resulting in 213,182 reviews. As the study focuses on breakdowns in user experience, we then applied a pretrained sentiment classifier from *TextBlob,* a widely used natural language processing library, to identify reviews labeled as negative (Steven Loria, 2018). This filtering step produced a corpus of 15,090 negative reviews, which forms the analytical foundation of this study. By focusing on negative evaluations, the dataset concentrates on moments where users explicitly articulate unmet expectations and perceived failures.

Prior to analysis, we conducted standard text preprocessing to prepare the corpus for computational modeling, including lowercasing texts, removing URLs and email addresses, and normalizing spaces. We tokenized the text at the word level and removed common English stop-words. The cleaned texts were then relinked to their original metadata to support subsequent quantitative and qualitative analyses. This preprocessing pipeline ensured that the corpus remained analytically tractable while preserving its connection to contextual attributes.

### Step 3: Analysis

To identify recurring patterns in user reviews without imposing predefined categories, we employed topic modeling as an inductive analytical approach. We trained a Latent Dirichlet Allocation (LDA) model on the corpus of 15,090 reviews (classified as negative using NLP), using a bag-of-words representation. LDA was selected because it offers an interpretable way to surface co-occurring terms and latent themes in large text corpora, providing a structured qualitative interpretation (Jelodar et al., 2019). Following iterative evaluation across a range of topic numbers (10–35), we selected a 10-topic solution based on the best coherence score (0.53) and the substantive interpretability of the resulting topics, with model parameters set to 100 iterations and 20 passes.

We treated the topic model as a heuristic tool to guide, rather than determine, our analysis. For each topic, we examined high-probability terms and reviewed a purposive sample of reviews with high topic proportions. Through comparative reading of these texts, we identified recurrent issues, evaluative language, and references to system behaviour, particularly in relation to health and AI interaction. Based on this process, we assigned each topic a descriptive label and a concise thematic summary. This interpretive step draws inspiration from thematic analysis principles in the sense articulated by Braun and Clarke (2006), using patterns suggested by the model as a starting point and refining them through close engagement with the data, rather than treating topics as fixed, purely statistical categories.

We then aggregated the 10 topics into three higher-level concern types that captured broader patterns across the dataset: **Access Barriers and Service Unreliability**, **User Experience and AI Interaction Quality**, and **Billing and Customer Support Issues**. Two authors collaboratively reviewed and refined these groupings, iteratively revisiting representative reviews to ensure conceptual alignment between topics and categories. This process positions computational modeling as a starting point for human-centered interpretation, allowing categories to emerge through engagement with the data rather than being imposed a priori.

### *Explicit Security Privacy mentions*

Explicit security, privacy, and data-handling mentions. In addition to topic-based analysis, we introduced a complementary lens to identify reviews that explicitly reference security-, privacy-, or data-handling-related concerns (SPR). We constructed a keyword-based lexicon (e.g., "privacy," "data sharing," "sell my data," "without consent") and, after normalising review texts, flagged any review containing at least one lexicon term as SPR-related (*spr_flag* = 1). This approach provides a cautious estimate that captures only reviews where such concerns are explicitly named. As a result, it likely underrepresents implicit worries and may include occasional false positives, so we treat SPR-related findings as exploratory and descriptive rather than definitive.

## RESULTS

We organize the results around the three higher-level concern categories derived from topic modeling. In the subsections that follow, we describe how each category is composed, and how users articulate concerns. Table 1 summarizes the distribution of reviews across the three categories. Across all three categories, mean ratings remain low, consistent with our focus on clearly negative evaluations. Figure 1 depicts the percentage distribution of the three concern categories by platform

| Category | Reviews | Platform Count | | Mean Rating (SD) | Median Rating | SPR Rate |
|---|---|---|---|---|---|---|
| | | Android | iOS | | | |
| Access Barriers & Service Unreliability | 3,197 | 2,922 | 275 | 1.713 (1.3) | 1.0 | 0.011 |
| Billing & Customer Support Issues | 2,775 | 2,407 | 368 | 1.548 (1.19) | 1.0 | 0.02 |
| User Experience & AI Interaction Quality | 9,118 | 7,874 | 1,244 | 2.875 (1.74) | 3.0 | 0.003 |

**Table 1. Summary of the three categories**

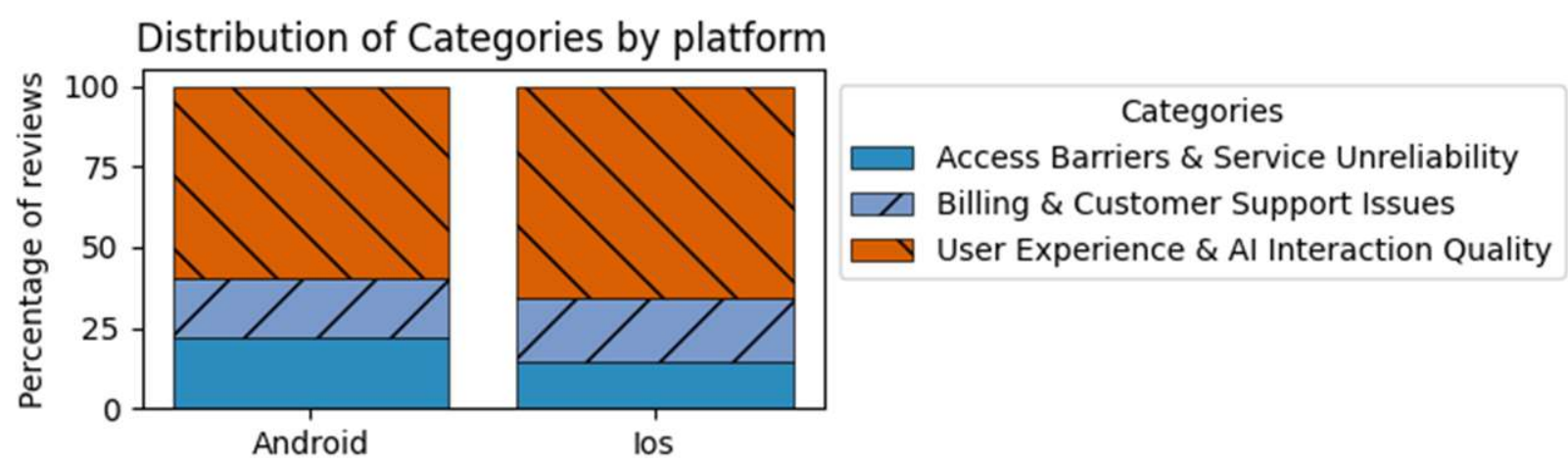

**Figure 1. Percentage distribution of the three concern categories by platform**

### Category 1: Access Barriers and Service Unreliability

This category captures breakdowns that prevent users from accessing or reliably using chatbots as functional information resources. It comprises three topics: **Access Barrier** ($n$ = 2,076; $M$ = 1.58), **Performance and Mental Health Issues** ($n$ = 230; $M$ = 2.27), and **UX Dissatisfaction** ($n$ = 891; $M$ = 1.89). Overall, the category shows consistently low ratings ($M$ = 1.71, $SD$ = 1.30; *median* = 1), indicating strong dissatisfaction and limited tolerance for failures at the point of access. These patterns suggest that for many users, breakdowns occur before meaningful interaction can take place.

The **Access Barrier** topic, the largest in this category, reflects two recurring forms of failure: restricted access through paywalls and technical unavailability. Many users report discovering, often during onboarding, that core features require payment, as in one review that states *"it was not free… you need to pay"* or another where a user describes answering all onboarding questions only to find *"it's not free"* and deciding to delete the app. Others emphasize basic login and connectivity problems that entirely block use, such as *"I already have an account but when I log in it does not work"* or *"there is a bug while login… I'm not able to log in and app is not responding"*. These accounts underscore how mismatches between expectations of free or low-cost access and actual subscription requirements, coupled with instability and failed authentication, prevent chatbots from functioning as dependable points of entry for health support.

The **Performance and Mental Health Issues** topic highlights a more consequential form of breakdown, where users link system failures to negative emotional outcomes. Some reviews describe the chatbot as unresponsive or irrelevant to the user's concerns, reporting that it *"just gives some random texts that are nowhere related to the concern you're speaking about"* and that *"instead of helping to improve your mental health, this app just makes it worse"*. Others criticize repetitive and non-learning behavior from both the chatbot and human coach, noting that the system *"keeps repeating the same stuff over and over again"* and that the coach *"kept giving generic advice"*. These responses illustrate how technical failures and limited adaptivity can exacerbate distress for users already in vulnerable states, rather than providing the intended support.

The **UX Dissatisfaction** topic reflects highly negative evaluations that often summarize broader frustrations rather than pinpointing a single issue. Reviews in this group use strong evaluative language to characterize the entire service, for example describing it as *"the worst app I have ever downloaded"* or *"one of the worst service provider[s]"* and warning others not to use it. Some users link this dissatisfaction to perceived low effort or generic content, such as a complaint that coaches *"will just copy and paste a internet pdf and ask you to follow"*, while others point to pervasive bugs and failures, calling it *"the worst app I had ever seen it haves more bugs"*. With the most negative sentiment polarity among all topics, this cluster captures reviews with particularly high affective intensity. Taken together, these topics point to a consistent pattern: when access is blocked, unstable, or experienced as deceptive, the chatbot fails not only as a technical system but as an information infrastructure that users can depend on in moments of need. Table 2 summarizes the three topics within this category.

| Topic Label | Reviews | | Dominant Complaint |
|---|---|---|---|
| | Count | Mean Rating (SD) | |
| Access barrier | 2076 | 1.58 (1.12) | Not free as expected, crashes, app stops working, comments removed |
| Performance and Mental Health Issues | 230 | 2.27 (1.44) | Crashes, slow, random or unrelated texts; sometimes worsens mental health |
| UX Dissatisfaction | 891 | 1.89 (1.56) | Strongly negative overall experience; 'worst app/service eve |

**Table 2. Topic breakdown of category 1**

### Category 2: User Experience and AI Interaction Quality

This category captures issues and concerns in how users experience and interact with chatbots during use. It includes five topics: **Interface and Design Issues** ($n$ = 437; $M$ = 2.24), **Perceived Uselessness** ($n$ = 1,261; $M$ = 1.62), **Poor Emotional Support** ($n$ = 3,435; $M$ = 4.06), **Lack of Responsiveness and Personalization** ($n$ = 1,759; $M$ = 2.45), and **Outdated and Low AI Agent Quality** ($n$ = 2,226; $M$ = 2.22). With 9,118 reviews (60.4% of the corpus), this is the largest category. It also has a higher mean rating ($M$ = 2.88, $SD$ = 1.74) than the other categories, suggesting a more varied range of dissatisfaction concerns. While many users report frustration, others assign moderate ratings, indicating partial rather than complete rejection of the system. Figure 2 visualizes these differences by showing the spread of ratings across all topics, highlighting both the low-rated themes and dispersed rating.

The **Poor Emotional Support** topic is the largest in the dataset and stands out for its relatively high mean rating compared to other themes. This pattern suggests that some users remain engaged enough to evaluate the chatbot's emotional value, even when dissatisfied. Reviews often describe the system as "*not that bad*" but ultimately unhelpful, confusing, or inconsistent in its responses. The sentiment polarity (−0.25) is less negative than other topics, reinforcing this interpretation. Rather than abrupt rejection, these reviews reflect unmet expectations in emotional support, where users seek meaningful engagement but encounter limited or superficial responses.

The topics **Perceived Uselessness** and **Outdated and Low AI Agent Quality** reflect a shared concern with the gap between expectations of AI capability and actual system performance. Users frequently report scripted interactions, repetitive responses, or features that resemble basic decision trees rather than adaptive conversational agents. For instance, one user characterizes the service as *"simply a waste of time and money… even after paying I don't get any personalized diet or workout"*, while another concludes that the app is *"mostly useless… basically a fasting timer and a very limited meal tracker"* and warns others *"don't waste your money"*. Complaints about outdated functionality and shifting feature sets further suggest that users evaluate these systems against rapidly evolving standards of AI performance, noting, for example, that *"the older version… was much better than this"*. In these cases, the chatbot fails as a source of relevant and personalized information, limiting its perceived usefulness.

The **Lack of Responsiveness and Personalization** topic highlights specific interactional concerns, where users report that the chatbot does not respond to their input in a meaningful way, instead defaulting to generic or mismatched replies (e.g., *"it doesn't seem to understand me at all"*). These interactions often lead to frustration and, in some accounts, increased emotional distress. The **Interface and Design Issues** topic adds a structural dimension, pointing to bad user experience, confusing navigation, and billing or cancellation processes that are intertwined with interface problems. Users describe the app as *"quite difficult to use"* and *"not intuitive"*, and some link interface confusion directly to unwanted charges, noting that they *"have been charged for membership and did not want to join or pay"*. Together, these topics show that even when access is technically possible, limitations in interaction quality and interface design constrain the chatbot's role as a responsive and supportive information system.

| Topic Label | Reviews | | Dominant Complaint |
|---|---|---|---|
| | Count | Mean Rating (SD) | |
| Interface & Design Issues | 437 | 2.245 (1.549) | Cancellation / account problems and poor interface / 'no real AI' |
| Lack of responsiveness & personalization | 1,759 | 2.454 (1.535) | Does not understand user; trigger-word driven, ignores actual problem |
| Outdated and Low AI Agent Quality | 2,226 | 2.216 (1.418) | Outdated or 'fake' AI agent; childish decision tree; tracking/logging frustrations |
| Perceived Uselessness | 1,261 | 1.621 (1.228) | Waste of time, wrong answers, low reliability |
| Poor Emotional Support | 3,435 | 4.059 (1.502) | Not helpful for emotional support; confusing or slow responses |

**Table 3. Topic breakdown of category 2**

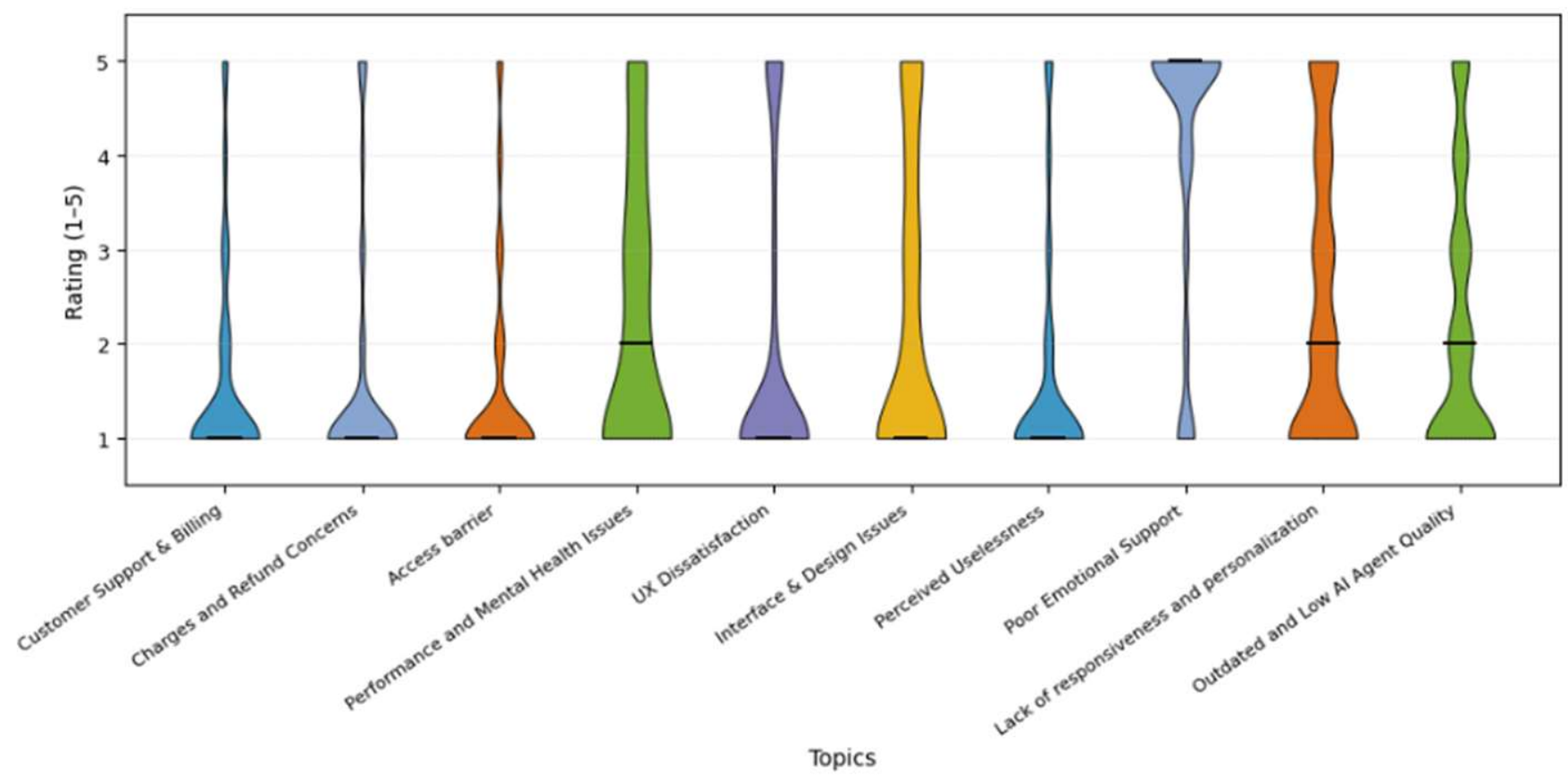

**Figure 2. Rating distributions across the topics**

### Category 3: Billing, Customer Support, and Trust

This category captures topics related to billing practices, customer support, trust and users' billing concerns. It covers two topics: **Customer Support and Billing** ($n$ = 689; $M$ = 1.53) and **Charges and Refund Concerns** ($n$ = 2,086; $M$ = 1.55). Although it is the smallest category by volume (2,775 reviews; 18.4% of the corpus), it shows the lowest mean rating ($M$ = 1.55; $SD$ = 1.19) and the most concentrated distribution of ratings, with both the median and 75th percentile at 1. This indicates a highly consistent and strong level of dissatisfaction compared to other categories. Table 4 summarises the two topics within this category.

The **Charges and Refund Concerns** topic focuses on unexpected charges, difficulties with cancellation, and unsuccessful refund attempts. Users frequently describe being billed in ways they perceive as deceptive, with one reviewer warning that there are *"hidden fees… on top of the very expensive subscription fees"* and concluding *"what a scam"*. Others emphasize that renewal charges arrive without clear warning, as in a review noting that there is *"no courtesy notification before you are charged… be wary of their trial period"*. Several accounts portray failed attempts to resolve these issues, for example a user who reports that the app *"auto charged for [a] smart plan… I complained and asked for a refund on day 1… they refused"*. These reviews point to a perceived lack of accountability and transparency in financial practices, undermining users' willingness to trust the service.

The **Customer Support and Billing** topic centers on perceived value and the responsiveness of human support. Users report that paid features do not meet expectations, describing the service as *"not worth the money"* and criticizing "*AI coaches*" who appear to provide generic or pre-packaged content rather than personalised guidance. For instance, one reviewer writes *"very unprofessional… not true experts at all … act as dietitians and sell pre-created plans with no personalisation"*. Others highlight difficulties in reaching effective support, stating that the company is *"available for calls when they want to sell any plans"* but that *"if you need any customer support, you have to raise chat and waste your time on nonsense"*. These accounts reinforce the perception that once payment is made, user concerns may not be adequately addressed.

Taken together, these patterns extend beyond routine consumer complaints. For users seeking health support, often as a lower-cost alternative to formal care, unexpected charges and unresolved billing issues represent an added burden. In such cases, the system fails both as a source of support and as a fair financial exchange, raising equity concerns about who can reliably access and benefit from AI healthcare chatbots. This category also has the highest proportion of security-privacy concerned reviews, suggesting that concerns about privacy and data practices can intersect with financial distrust in contexts where users already feel misled or poorly supported.

| Topic Label | Reviews | | Dominant Complaint |
|---|---|---|---|
| | Count | Mean Rating (SD) | |
| Charges and Refund Concerns | 2086 | 1.555 (1.232) | Unexpected charges and difficulty obtaining refunds |
| Customer Support & Billing | 689 | 1.527 (1.065) | Billing / paid features & value complaints |

**Table 4. Topic breakdown of category 3**

## SPR complaints

This section examines reviews that explicitly reference privacy, security, or data-handling concerns as a distinct dimension of user dissatisfaction. Using the keyword-based procedure described in the Method section, we identified 118 reviews as SPR-related in the negative-review corpus. These reviews appear most frequently within the Billing and Customer Support Issues category, followed by Access Barriers and Service Unreliability, and least often in User Experience and AI Interaction Quality. As discussed earlier, this distribution suggests that explicit SPR concerns tend to arise alongside problems of billing and financial trust.

Across these reviews, users often frame SPR issues as breaches of trust rather than as minor inconveniences. Some accounts characterize the entire service as deceptive or hostile to privacy, describing it as *"worthless smoke and mirrors hiding behind a veil of 'privacy'"* or warning that the company will *"take your very personal details and not only overstep the privacy boundaries but you'll be spammed out for life… install at your own risk."* Others object to what they see as unnecessary or opaque data collection at the point of registration, noting that the app demands *"unnecessary personal data collection… you have to sign up with your email, pointless data collection"* or concluding that they *"would never subscribe to a service that has no function and disrespects personal privacy."* These complaints position privacy violations as a reason to reject the service outright.

Users also question how their data are stored and shared once collected. One reviewer describes an *"invasive privacy policy … sells your identity and health info to other companies,"* while another expresses unease upon learning that *"everything is logged"* and that a developer could access their message history, asking whether it is acceptable that *"a dev can access your private message log at any time with the press of a few buttons."* In such accounts, concerns about logging, third-party cookies, and data sharing are closely tied to broader feelings of surveillance and loss of control, even when the nominal privacy policy promises protections.

SPR-flagged reviews also show substantially lower ratings than non-flagged reviews ($M$ = 1.23, $SD$ = 0.74 as compared to M = 2.39, $SD$ = 1.68; $t \approx 16.67$, $p < .001$), as shown in Figure 3. Although relatively few, reflecting the cautious nature of keyword-based identification, these reviews cluster among the most negative evaluations in the dataset. This pattern indicates that when users explicitly raise privacy, security, or data concerns, they do so in contexts of heightened dissatisfaction and distrust rather than routine critique.

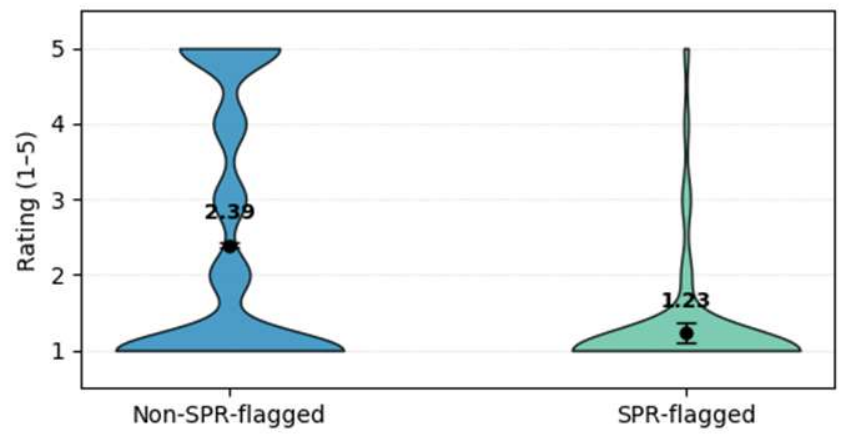

**Figure 3. Rating distributions for SPR vs. non-flagged reviews**

## DISCUSSION

### Interpreting Negative Experiences as Infrastructural User Concerns

The patterns identified in this study point to recurring issues in how AI healthcare chatbots function as information infrastructures in everyday use (Bowker et al., 1996; Wah, 2025). Rather than isolated usability problems, the three concern types reflect failures at different stages of users' engagement with these systems. Billing and customer support issues indicate dissatisfaction with the billing, transparency and customer support that govern access, echoing prior work on the centrality of trust and accountability in digital health services. Access barriers and service unreliability reflect failures at the point of entry, where users are unable to reach or sustain interaction with the system, a pattern that resonates with studies showing how infrastructural fragility can keep formal services "out of reach" even when they are nominally available. In turn, interaction-related concerns arise when users can engage but find the system unresponsive, unhelpful, or misaligned with their needs, extending earlier findings on limited accuracy and inconsistent conversational behavior in health chatbots by demonstrating how such limitations are experienced at scale in everyday settings.

While paywalls, crashes, and billing disputes are not unique to AI healthcare chatbots, they acquire a distinctive character in this context. Users of general productivity or entertainment apps who encounter paywall are denied convenience; users of mental health chatbots who encounter the same barrier may be denied access to emotional support at a moment of distress (Zhang et al., 2024; Vaidyam et al., 2019; Torous et al., 2023). Similarly, AI-specific failures, scripted dialogue masquerading as adaptive conversation, outdated agents that do not respond to user input, and interactions that users explicitly describe as "not real AI", reflect a gap between the affordances users associate with modern generative AI and the actual capabilities of the systems they encounter (Zhang et al., 2024, Miner et al., 2016; Fitzpatrick et al., 2017). These are not generic mobile app problems; they reflect unmet expectations specific to conversational AI in health-sensitive contexts.

These breakdowns also differ in how users experience and evaluate them. Failures related to billing and access tend to produce immediate and uniformly negative responses, as they prevent meaningful use altogether, aligning with prior research that links to lack of financial and billing transparency and restricted access to sharp declines in trust. By contrast, interaction-related issues show greater variation: some users disengage quickly, while others continue interacting long enough to articulate specific limitations, such as lack of personalization or shallow emotional support. The relatively higher ratings observed in parts of this category suggest that dissatisfaction can coexist with partial engagement, particularly when users approach these systems as ongoing companions rather than one-off tools. This nuance adds to previous work by showing that negative experiences are not uniform; some signal complete failure, while others reveal tensions within continued reliance.

Viewed together, these findings frame AI healthcare chatbots as infrastructures that often fail across access, interaction, and governance dimensions. This perspective shifts attention beyond technical performance to the broader conditions under which such systems become usable, meaningful, and trustworthy in practice. In doing so, the study complements controlled evaluations of chatbot accuracy and safety by stressing how users themselves describe breakdowns in trust, access, and everyday interaction, highlighting areas where future work on design, regulation, and policy must be attentive to infrastructural as well as interface-level concerns.

### Implications for Well-Being and Information Practice

These infrastructural breakdowns carry particular significance in the context of health support. Users of these systems are often seeking assistance for mental health support such as anxiety, stress, or emotional distress. In this context, failures in access, interaction, or billing are not merely technical inconveniences. Instead, they can interrupt or undermine attempts to seek support, sometimes at moments of heightened vulnerability (Vaidyam et al., 2019; Haque & Rubya, 2023). User accounts describing frustration, confusion, or worsening emotional states point to the potential consequences of unreliable or poorly aligned systems in sensitive-use contexts.

From an information science perspective, these findings reinforce the importance of access, usability, and trust as foundational conditions for meaningful information use. Prior work has emphasized that equitable access to information systems is necessary for users to benefit from them (Bowker & Star, 1999; Choi & DiNitto, 2013;Jaeger et al., 2007). The results here extend this insight by showing that access alone is insufficient. Systems must also be intelligible in interaction and fair and transparent in their financial and billing process and customer support arrangements (Kaihlanen, 2022; Torous et al., 2023). When these conditions are not met, the informational value of the system is diminished, regardless of its underlying technical capabilities.

For designers and policymakers, the results highlight areas where improvements are both necessary and actionable. Issues such as unclear subscription models, limited support channels, and opaque billing practices are not peripheral concerns; they shape whether users can access and trust these systems at all. Addressing these challenges requires attention not only to interface design or algorithmic performance, but also to transparency, accountability, and regulatory oversight. There is a need to align consumer-facing AI healthcare applications with existing and emerging digital-health regulations, including clearer standards around consent, subscription renewal, and disclosure of data uses. This is especially important given that many consumer health applications operate outside traditional healthcare governance frameworks, such as HIPAA in the United States (Marks & Haupt, 2023; Hassan & Bashir, 2023). Strengthening regulatory guidance on billing transparency, data use, and redress mechanisms would help ensure that such systems function as equitable and reliable components of digital health environments.

### Data Governance Practices as Conditional Signals of Distrust

The analysis of SPR concerns adds a more focused perspective on how users articulate distrust. Explicit mentions of SPR issues are relatively limited in the corpus, yet they are associated with the most negative ratings. This pattern suggests that privacy and data concerns are not routinely expressed, but when they do surface, they signal particularly strong dissatisfaction.

One interpretation is that SPR concerns require a different kind of user reasoning. While access problems, poor interaction, or unexpected charges are immediately visible, concerns about data handling often involve less observable processes, such as third-party data sharing, cross-app tracking, or long-term storage. Prior work suggests that users do not actively engage with these issues unless prompted by a negative experience or a breakdown in trust (Ischen et al., 2020; Gumusel, 2025). The findings here are consistent with this view, as SPR concerns tend to appear alongside other forms of dissatisfaction rather than in isolation.

The concentration of SPR-related reviews within the billing and customer support category further supports this interpretation. Experiences such as unexpected charges or unresolved billing issues may prompt users to question broader aspects of the system, including how their data are managed. In contemporary mobile ecosystems, these worries are shaped not only by the apps themselves but also by platform-level data-governance features and their limitations, for example, privacy "*nutrition labels*" on iOS or permission dashboards on Android that promise transparency but may still leave users uncertain about what is collected, how long it is retained, and with whom it is shared. Cases where users feel misinformed or misled, such as discovering extensive logging only after probing a chatbot about its data practices, illustrate how gaps between stated policies and perceived behavior can erode trust. In this sense, privacy and security concerns may emerge not as primary points of user dissatisfaction, but as secondary signals that reflect deeper issues of trust (e.g., Sharma & Bashir, 2020; Yener et al., 2025). Strengthening data-governance practices, through clearer privacy notices, more granular consent mechanisms, and independently verifiable disclosures about data flows, may therefore have indirect benefits for trust, particularly when coupled with fair billing and responsive support. Taken together, these findings position SPR concerns as an important but conditional dimension of user experience. They highlight the need to consider privacy and data practices not only as abstract principles, but as elements that become salient where systems fail to meet user expectations more broadly.

### Limitations and Future Work

This study relies on self-reported app-store reviews, which may not fully represent broader user populations, and on interpretive analytical choices (e.g., topic modeling, category aggregation) that could yield slightly different thematic structures under alternative specifications. The keyword-based search only captures only explicit mentions of SPR concerns, potentially missing more implicit forms of discomfort, and the analysis does not account for linguistic or demographic variation that may shape how users experience these systems. Future work could combine large-scale review analysis with qualitative methods (interviews or surveys) to link observed patterns more directly to user well-being and trust, conduct comparative analyses across regions, languages, or health domains, and develop more context-sensitive approaches for detecting SPR concerns to inform design and policy interventions.

## CONCLUSION

This study examined user reviews of AI healthcare chatbots through large-scale analysis of app-store reviews. We showed how users describe barriers to accessing and reliably using these systems, highlighting paywalls, instability, and login failures that prevent interaction altogether. We also investigated how users characterise interaction quality and perceived usefulness, identifying frustrations with limited emotional support, scripted or "fake" AI behaviour, and interfaces that make ongoing use difficult even when access is available. Finally, we documented how users experience trust, fairness, support structures, and explicit privacy/security concerns, with billing disputes and weak customer support eroding confidence in the service and a small but important subset of reviews expressing strong worries about data collection, tracking, and misuse.

Taken together, these findings frame AI healthcare chatbots as information infrastructures whose value depends on the alignment of access conditions, interaction quality, and governance practices. App-store reviews emerge as a useful lens on these dynamics, capturing how systems are evaluated in everyday contexts rather than only in controlled settings. As AI-enabled healthcare tools continue to evolve, ensuring that they function as reliable, equitable, and trustworthy information systems will require attention not only to conversational performance, but also to billing transparency, support processes, and robust, comprehensible data-governance practices.

## GENERATIVE AI USE

All work, analysis, and writing were conducted by the author(s), who take full responsibility for the content. Generative AI was used only for minor grammatical edits.